\documentclass[twocolumn,nofootinbib,preprintnumbers,amsmath,amssymb]{revtex4}

\usepackage{graphicx}
\usepackage{dcolumn}
\usepackage{bm}
\usepackage{amsmath}

\def \be {\begin{equation}}
\def \ee {\end{equation}}

\begin{document}

\title{How Pulsars Shine II: TeV emission}

\author{Andrei Gruzinov}

\affiliation{ CCPP, Physics Department, New York University, 726 Broadway, New York, NY 10003
}

\date{October 18, 2023}

\begin{abstract}

Recent discovery of 20 TeV radiation from the Vela pulsar confirms (tentatively, at the level of crude estimates) the Aristotelian Electrodynamics picture of pulsar radiation: pulsars shine, mostly in GeV, by annihilating colliding Poynting fluxes into curvature radiation near the light cylinder. The observed GeV photons are the curvature radiation of electrons/positrons with  Lorentz factors $\sim 10^7-10^8$. These ``super-ultra-relativistic'' electrons/positrons must also produce TeV radiation by inverse Compton if low-energy target photons are available, as they are in the Vela pulsar. 

\end{abstract}

\maketitle

\section{Introduction}

We first describe a scenario for the Vela pulsar's TeV radiation, \S\ref{S:tev}. This scenario is based on Aristotelian Electrodynamics (AE, \S\ref{S:ae}) as applied to weak, Geminga-like, pulsars, \S\ref{S:w} ({\it weak} pulsars do not produce plasma near the light cylinder). The Vela pulsar is {\it strong}, it does produce plasma near the light cylinder, \S\ref{S:s}. AE still applies, but it must be supplemented by a pair production description. We will not describe the pair production explicitly. We will combine some of the observables and the AE pulsar theory to explain TeV (\S\ref{S:tev}). 20 TeV gamma rays from the Vela pulsar \cite{HESS} confirm the AE pulsar theory. Only tentatively, of course. 

There are no new results in this paper. We merely summarize the many previous papers to bring out the simplicity of the AE picture of pulsar radiation: 
\begin{itemize}
\item As described by Goldreich and Julian \cite{GJ} even axisymmetric pulsars spin down, because they create plasma. The plasma supports a large-scale electromagnetic field with an outward Poynting flux component. 
\item Guided by the poloidal magnetic field of the neutron star, a fraction of the Poynting flux coming from the magnetic North collides, near the light cylinder, with a fraction of the Poynting flux coming from the magnetic South. 
\item Goldreich and Julian \cite{GJ} postulated that the colliding Poynting fluxes will repel and escape to infinity, without any loss of energy. In reality, that is in AE, colliding Poynting fluxes annihilate into curvature radiation. 
\item We know that pulsars radiate in the AE regime because the observed GeV cutoffs are well above $\frac{mc^2}{\alpha}=70$MeV.
\item The AE pulsar theory works well for weak, Geminga-like, pulsars. For strong, Vela-like, pulsars, soft X-rays from the star and the curvature GeV gamma rays produce electron-positron pairs near the light cylinder. This reduces the GeV cutoff energy and the pulsar efficiency, but curvature radiation of electrons/positrons with  Lorentz factors $\sim 10^7-10^8$ remains the primary radiation mechanism (as tentatively confirmed by the Vela's pulsed TeV).
\end{itemize}

\section{TeV radiation of the Vela pulsar}\label{S:tev}

We first list all the observational input and the derived data needed for the analysis in this section and in \S\ref{S:s}. Then we describe a scenario for the Vela pulsar's TeV radiation.

\subsection{Data}

We need: pulsar parameters, GeV radiation, TeV radiation, high-frequency radio, neutron star temperature. From [1-4], without error bars, as only crude estimates will follow:
\begin{itemize}
\item distance to Vela $d=0.29$kpc;
\item pulsar period $P=89$ms;
\item pulsar spindown rate $\dot{P}=1.25\times 10^{-13}$;
\item GeV energy flux $f_{\rm GeV}=9.1\times 10^{-9}\frac{\rm erg}{{\rm cm}^2\cdot {\rm s}}$ with cutoff energy $E_{\rm cut}=3.0$GeV;
\item TeV energy flux $f_{\rm TeV}\sim 10^{-13}\frac{\rm erg}{{\rm cm}^2\cdot {\rm s}}$ at $\sim 10$TeV;
\item Pulsar radio emission $F_\nu=67\mu$Jy at $\nu=344$GHz, spectral index $\alpha=-0.93$;
\item $T_{\star}=0.66$MK.
\end{itemize}

Derived parameters:
\begin{itemize}
\item GeV luminosity $L_{\rm GeV}=9.2\times 10^{34}\frac{\rm erg}{\rm s}$;
\item TeV luminosity $L_{\rm TeV}\sim 10^{30}\frac{\rm erg}{\rm s}$;
\item light cylinder radius $R=4.3\times 10^8$cm;
\item characteristic electromagnetic field in the light cylinder region  $B=36$kG;
\end{itemize}
In the above: $L=4\pi d^2f$, $R=\frac{cP}{2\pi}$, $cB^2R^2=\dot{E}=\frac{4\pi^2I\dot{P}}{P^3}$, $I=10^{45}{\rm g}\cdot{\rm cm}^2$.

\subsection{AE Interpretation}

According to the AE picture of pulsars (\S\ref{S:w}, \S\ref{S:s}), the $E_{\rm cut}=3.0$GeV photons are emitted by electrons (from now on ``electrons'' means electrons and/or positrons) moving along trajectories with radius of curvature $\sim R$, at the peak of the curvature radiation power:
\be
E_{\rm cut}\sim 0.29\times \frac{3}{2} \gamma^3 \frac{c\hbar }{R}.
\ee
(Here and below, we use ``$\sim$'' while actually calculating to 2 digits. We must use ``$\sim$'' because our assumptions, like radius of curvature $R$ instead of $2R$, are meaningful only to order of magnitude. We calculate to 2 digits to avoid error accumulation.) This gives the Lorentz factor $\gamma$ and energy $E_e$ of the GeV-emitting electrons
\be
\gamma \sim 5.3\times 10^7,~~~E_e=27{\rm TeV}.
\ee

\subsection{TeV radiation}
These super-ultra-relativistic electrons will scatter a target photon of energy $E_t\sim \frac{mc^2}{\gamma}\sim 9.6\times 10^{-3}$eV into a photon of energy $E_{\rm IC}\sim \frac{2}{3}E_e\sim 18$TeV with the cross section $\sigma_{\rm IC} \sim 0.47 \times \frac{8\pi}{3}r_e^2\sim 3.1\times 10^{-25}{\rm cm}^2$. 

Photons of energy $E_t\sim \frac{mc^2}{\gamma}\sim 9.6\times 10^{-3}$eV have frequency $\nu \sim 2.3$THz. Extrapolating from $F_\nu=67\mu$Jy at $\nu=344$GHz with the spectral index $\alpha=-0.93$, we get $F_{\nu t}=11\mu$Jy at $\nu _t=2.3$THz. The density of the target photons near the light cylinder is 
\be
n_t\sim \frac{F_{\nu t}}{2\pi c\hbar}\left(\frac{d}{R}\right)^2\sim 2.4\times 10^{12}{\rm cm}^{-3}.
\ee
The optical depth for Compton scattering is 
\be
\tau \sim n_t\sigma_{\rm IC}R\sim 3.2\times 10^{-4} . 
\ee

The number of electrons responsible for the observed gamma radiation is 
\be
N\sim \frac{L_{\rm GeV}}{\frac{2}{3}\gamma^4\frac{ce^2}{R^2}}\sim 4.7\times 10^{29}.
\ee
The Inverse Compton luminosity of these electrons is
\be
L_{\rm TeV, theoretical}\sim \frac{N\tau E_{\rm IC}}{R/c}\sim 3.0\times 10^{29}\frac{\rm erg}{\rm s},
\ee
versus the observed $L_{\rm TeV}\sim 10^{30}\frac{\rm erg}{\rm s}$ -- acceptable.

The above might all be very well, but not particularly interesting unless we explain where the (i) $N\sim 10^{30}$ electrons, (ii) with Lorentz factors $\gamma \sim 10^8$, and (iii) radii of curvature $R\sim 3\times 10^8$cm come from. 

We will compute these three numbers:
\begin{itemize}
\item {\it exactly} for a ``pulsar device'', \S\ref{S:ae};
\item {\it convincingly}, although numerically, for a weak pulsar, \S\ref{S:w}; 
\item {\it tentatively}, very crudely, for a strong pulsar, \S\ref{S:s}.
\end{itemize}

\section {Aristotelian Electrodynamics}\label{S:ae}

To the best of my knowledge, AE first appeared in \cite{Fink}. 

\subsection{AE: definition and estimates}

Aristotelian Electrodynamics (AE) \cite{Fink, Gruz, Jaco} is a simplified description of electron/positron motion {\it and curvature radiation} in a strong electromagnetic field. AE is valid when electrons are radiation-overdamped; this happens when the Aristotle number (analog of the inverse Reynolds number -- the dynamical time divided by the damping time, to be explained) is large:
\be
{\rm Ar}\equiv \frac{e^{5/4}}{mc^2}B^{3/4}R^{1/2}=\left(\frac{B}{B_c}\right)^{3/4}\left(\frac{R}{r_e}\right)^{1/2}\gg 1.
\ee
Here $r_e\equiv\frac{e^2}{mc^2}=2.82\times 10^{-13}$cm is the classical electron radius, $B_c\equiv \frac{e}{r_e^2}=6.04\times 10^{15}$G is the classical electron field, $B$ is the characteristic electromagnetic field, $R$ is the characteristic length scale. 

In the above, one assumes that the work done by the electric field $E\sim B$ is balanced by curvature radiation:
\be
ceB\sim \gamma^4\frac{ce^2}{R^2}.
\ee
This gives the terminal Lorentz factor
\be
\gamma\sim \left(\frac{BR^2}{e}\right)^{1/4}.
\ee
The ratio of the dynamical time and the curvature cooling time is called Ar:
\be
{\rm Ar}\sim \frac{R/c}{\gamma mc^2/\frac{\gamma^4ce^2}{R^2}}.
\ee

The overdamped electrons emit curvature gamma rays of characteristic energy (critical energy of the synchrotron spectrum)
\be\label{ecar}
E_c\sim \frac{mc^2}{\alpha}\times {\rm Ar}=70{\rm MeV}\times {\rm Ar},
\ee
$\alpha\equiv \frac{e^2}{c\hbar}=\frac{1}{137}$ is the fine structure constant, because the energy of curvature photons is
\be
E_c\sim \gamma^3\frac{\hbar c}{R}.
\ee

Eq.(\ref{ecar}) immediately shows that pulsars must operate in the AE regime, as the observed GeV cutoff energies are above about 1 GeV; assuming the observed GeV radiation is curvature.

\subsection{AE: precise formulation}

Under the AE assumption, ${\rm Ar}\gg 1$, the velocity of positrons and electrons is equal to the speed of light: 
\be\label{ae}
{\bf v}_{\pm}=\frac{{\bf E}\times {\bf B}\pm(B_0{\bf B}+E_0{\bf E})}{B^2+E_0^2}~~~({\rm AE~I}).
\ee
This is the first AE equation. It says how the charges move. Here $E_0$ is the proper electric field scalar and $B_0$ is the proper magnetic field pseudoscalar:
\be\label{prop}
B_0^2-E_0^2=B^2-E^2,~ B_0E_0={\bf B}\cdot {\bf E},~ E_0\geq 0.
\ee
To derive the first AE equation, construct a future directed null 4-vector $(1,{\bf v})$ in terms of the electromagnetic field tensor $F_{\mu \nu}=({\bf E},{\bf B})$, say as an eigenvector of $F_{\mu \nu}$ \cite{Jaco}. 

The actual (very large) Lorentz factor is obtained by balancing the curvature power ($R$ is the radius of curvature)
\be 
q=\frac{2}{3}\gamma^4\frac{ce^2}{R^2},
\ee
by the electric field power
\be 
q=\pm e c{\bf v}_\pm\cdot{\bf E},
\ee
or, using Eq.(\ref{ae}),
\be 
q=ecE_0~~~({\rm AE~II}).
\ee
This is the second AE equation. It says how much power the charge radiates. All the power $q$ is emitted strictly along ${\bf v}_\pm$, as opening angles $\sim \gamma^{-1}\sim 10^{-8}$ don't matter.
Knowing $\gamma$, one computes the spectrum -- synchrotron with critical energy
\be\label{ecrit}
E_c=(3/2)^{7/4}c\hbar e^{-3/4}E_0^{3/4}R^{1/2}~~~({\rm AE~III}).
\ee
This is the third and last AE equation. It says with what spectrum the power $q$ is emitted.

\subsection{AE: exact solution -- the pulsar device}

The first AE equation, Eq.(\ref{ae}), plus Maxwell equations for the resulting electromagnetic field have the following exact solution called the pulsar device (in dimensionless rationalized units):
\be 
{\bf E}=(0,e^{-y},0),
\ee
\be
{\bf B}=(0,0,B(x,y)), 
\ee
\be 
B(x,y) = -e^{-y}\left\{ \begin{array}{rl}
{\rm sign}(x), & ~~~|x|>\frac{\pi}{2}; \\
\sin (x), & ~~~|x|<\frac{\pi}{2}. \\
\end{array}\right.
\ee
There are no positrons; electrons' number density $n$ and velocity ${\bf v}$ follow from $\nabla \cdot {\bf E}=-n$ and $\nabla \times {\bf B}=-n{\bf v}$:
\be
n=e^{-y},
\ee
\be 
{\bf v}= \left\{ \begin{array}{rl}
(-{\rm sign}(x),0,0)~~~~~~~~, & ~~~|x|>\frac{\pi}{2}; \\
(-\sin (x), -\cos(x),0), & ~~~|x|<\frac{\pi}{2}. \\
\end{array}\right.
\ee

Maxwell equations are thus satisfied. As a consequence, the continuity equation, $\nabla \cdot(n{\bf v})=0$, is satisfied too. It only remains to be checked that ${\bf v}={\bf v}_-$ as given by Eq.(\ref{ae}). 

First compute the proper fields:
\begin{eqnarray}
|x|>\frac{\pi}{2}:~~E_0=0,~~B_0=0~~~~~~~~~~~~~~; \\
|x|<\frac{\pi}{2}:~~E_0=e^{-y}\cos(x),~~B_0=0~~~.
\end{eqnarray}
We see that $|x|>\frac{\pi}{2}$ is a force-free zone, where no electromagnetic energy is transferred to electrons, while $|x|<\frac{\pi}{2}$ is a radiation zone, where the electromagnetic energy is transferred to electrons and radiated away. 

In the force-free zone the electrons move along $x$, toward $x=0$, that is toward the radiation zone. This is a pure $E\times B$ drift, at the speed of light, as dictated by Eq.(\ref{ae}). The electrons carry (more precisely, are accompanied by) a Poynting flux ${\bf S}=(-e^{-2y}{\rm sign}(x),0,0)$. The divergence of the Poynting flux is zero, $\nabla \cdot {\bf S}=0$. The entire Poynting flux coming from $x=\pm \infty$ reaches the radiation zone.

In the radiation zone the proper electric field $E_0$ appears and the electrons are deflected toward negative $y$, still as dictated by Eq.(\ref{ae}):
\be
{\bf v}_-=\frac{{\bf E}\times {\bf B}-E_0{\bf E}}{B^2+E_0^2}={\bf v}.
\ee
Here, in the radiation zone, at $|x|<\frac{\pi}{2}$, the Poynting flux is ${\bf S}=(-e^{-2y}\sin(x),0,0)$. The divergence of the Poynting flux is negative, $\nabla \cdot {\bf S}=-e^{-2y}\cos(x)$, there is an energy sink. None of the Poynting flux reaches the midplane $x=0$. The Poynting flux is completely annihilated into curvature radiation. 

\section{Weak pulsar}\label{S:w}

Essentially, a weak pulsar is a pulsar device with the size equal to the light cylinder radius and the characteristic EM field equal to the pulsar's field at the light cylinder.

Even the lowest spindown luminosity pulsars seen in gamma rays, like the Geminga pulsar, have ${\rm Ar}\gtrsim 10$ near the light cylinder. AE applies and can be used to describe pulsar radiation. This has been done in a long series of papers \cite{Gruzon}. The results are in good quantitative and/or qualitative agreement with the Fermi Gamma-ray Space Telescope pulsar data \cite{Fermi}, but only for {\it weak} pulsars, with moderate Aristotle numbers, ${\rm Ar}\lesssim 30$. Here we summarize the methods and results of \cite{Gruzon}. 

Methodologically, \cite{Gruzon} is straightforward. Take a small star -- small, compared to the light cylinder $R$. The star is a conducting rotating sphere, with a magnetic dipole moment. Electrons and positrons are copiously deposited near the star -- copiously, compared to the Goldreich-Julian number per rotation, $\dot{N}_{\rm GJ}\sim \frac{cBR}{e}$. Then the electrons and positrons move and radiate as dictated by AE. The electric and magnetic fields are given by Maxwell equations with appropriate boundary conditions at the surface of the star, with the current density from the AE-moving electrons and positrons. 

The results are qualitatively similar to the pulsar device with the size and field equal to the pulsar's actual $R$ and $B$ -- the light cylinder radius and the magnetic field at the light cylinder radius. The results are approximately independent of the size of the star -- so long as the star remains within the light cylinder radius $R$. The results are approximately independent of the pair deposition rate -- so long as the deposition rate stays above the Goldreich-Julian rate $\dot{N}_{\rm GJ}$. 

We get reasonable: lightcurves, characteristic (averaged over spin-dipole and observation angles) efficiency of the spindown to GeV conversion,
\be
\epsilon\equiv \frac{L_{\rm GeV}}{\dot{E}}\sim 0.2,
\ee
photon index 
\be
\Gamma \approx 1,
\ee
photon energy cutoff
\be\label{ecut}
E_{\rm cut}\approx 5.2\dot{E}_{34}^{3/8}P_{\rm ms}^{-1/4}{\rm GeV}.
\ee

The above is indeed very similar to the pulsar device. $E_{\rm cut}$ is a numerically computed version of Eq.(\ref{ecrit}). $\epsilon \sim 0.2$ rather than $\epsilon =1$ is because in a pulsar most of the Poynting flux escapes to infinity without colliding. Only a fraction of the Poynting flux, guided by the poloidal magnetic field, has to collide and annihilate near the light cylinder. 

It is clear why the vicinity of the light cylinder is (seen in the numerical simulations of \cite{Gruzon} to be) the main region for the Poynting flux collision. Inside the light cylinder, the poloidal magnetic field dominates, most of the plasma simply co-rotates with the star. Outside the light cylinder, the poloidal magnetic field becomes weaker that the radial Poynting flux. Far outside the light cylinder, the poloidal field cannot guide the Poynting flux, here the Poynting flux guides the poloidal field and escapes to infinity. 

So the Poynting fluxes mostly collide near the light cylinder. In order to mimic the pulsar, the size of the pulsar device must be $\sim R$, the light cylinder radius, and the field of the pulsar device must be $\sim B$, the pulsar field at the light cylinder. 

The synchrotron photon index is $\Gamma=2/3$, but the PLEC fit to a synchrotron spectrum that we found numerically for an axisymmetric pulsar is $\Gamma\approx 1$, clearly due to smeared $E_c$. Further increasing $\Gamma$ is not a problem  (and therefore not a good probe). But the observed $\Gamma>2/3$ is significant.

One might think that a fatal flaw of our approach is the large size of the star. Even in the best simulations we had $R_\star\gtrsim 0.1R$, while the actual size is $R_\star\sim 0.001R$. But actually our size is exactly right. If the plasma didn't exist at radii $\sim 0.1R$, the electric field would have been unscreened, $E_0$ would drive a test electron to high terminal Lorentz factors, the resulting GeV curvature photons would pair produce on the stellar soft X-rays, leading to an avalanche, similar to what happens in {\it strong} pulsars described in \S{\ref{S:s}}.

\section{Strong pulsar}\label{S:s}

For a weak axisymmetric pulsar, the GeV cutoff is given by Eq.(\ref{ecut}). For example, for the Geminga pulsar, with $\dot{E}=3.3\times 10^{34}\frac{\rm erg}{\rm s}$, $P=237$ms, we get $E_{\rm cut}=2.1$GeV from Eq.(\ref{ecut}) versus the observed $2.2\pm 0.1$ GeV. But Eq.(\ref{ecut}) clearly fails for the Vela pulsar: predicted $E_{\rm cut}=20$GeV versus the observed $3.0\pm 0.1$ GeV. This is because at high Aristotle numbers pair production at the light cylinder is inevitable, the pulsar becomes {\it strong}. 

The Aristotle number, in terms of the pulsar period $P$ and the spindown power $\dot{E}$ is
\be
{\rm Ar}=39\times\dot{E}_{34}^{3/8}P_{\rm ms}^{-1/4},
\ee
which is 16 for the Geminga and 147 for the Vela. 

Pair production by GeV gamma-rays on soft X-ray photons from a star of temperature $T_\star$ occurs if
\be
E_c\times T_\star\gtrsim (mc^2)^2,~~~E_c\sim \frac{mc^2}{\alpha}\times {\rm Ar}.
\ee
That is for ${\rm Ar}\gtrsim {\rm Ar}_c$, where 
\be
{\rm Ar}_c\sim \frac{\alpha mc^2}{T_\star}\sim \frac{3.7{\rm keV}}{T_\star}.
\ee
For the Geminga ($T_\star \approx 42$eV \cite{Karg}), ${\rm Ar}_c\sim 88>16$. For the Vela, $T_\star \approx 57$eV, ${\rm Ar}_c\sim 65<147$. 

We must conclude that, unlike the Geminga, the Vela would pair produce if it were in the {\it weak} pulsar regime. When the pair production threshold is exceeded, an avalanche of pair production occurs if the mean number of pairs produced by an electron as it crosses the light cylinder is large, $N_{\rm pair}\gtrsim 1$. Now 
\be
N_{\rm pair}\sim \tau _{\gamma X}N_{\rm GeV},
\ee
where $N_{\rm GeV}$ is the number of gamma-ray photons emitted by an electron as it crosses the light cylinder and $\tau _{\gamma X}$ is the optical depth for pair production.

We have
\be
N_{\rm GeV}\sim \frac{eBR}{E_c},
\ee
and 
\be
\tau _{\gamma X}\sim \sigma_{\rm pf}Rn_X,
\ee
\be 
\sigma_{\rm pf}\sim r_e^2\frac{m^2c^4}{E_cT_\star}\ln\left(\frac{E_cT_\star}{m^2c^4}\right),
\ee
\be
n_X\sim \frac{\pi}{240}\left(\frac{R_\star}{R}\right)^2\left(\frac{T_\star}{c\hbar}\right)^3,
\ee
giving $N_{\rm pair}\sim 14$ for the Vela. 

Both conditions are satisfied: we are above the threshold, ${\rm Ar}\gtrsim {\rm Ar}_c$, and strongly in the avalanche regime, $N_{\rm pair}\gg 1$. Pair production must occur near the light cylinder of the Vela pulsar. 

We know, tentatively \cite{Gruzs}, that pair production reduces (i) the size of the radiation zone, (ii) the proper electric field $E_0$.  For example, we have $E_{0~{\rm max}}=0.65B$ in AE without pair production (Fig.2 in the ``Pulsar Emission Spectrum'' paper of \cite{Gruzon}), and $E_{0~{\rm max}}=0.20B$ in a model AE with pair production (Fig.5 of the second paper of \cite{Gruzs}). 

The pair production avalanche stabilizes as the size and the strength of the radiation zone is reduced. But the controlled pair production never stops -- it is needed to maintain small $E_0$ and therefore small $E_{\rm cut}$. To get to the observed $E_{\rm cut}\sim 3GeV$ from the weak-pulsar theoretical value $\sim 20$GeV, we need roughly a factor of 10 ``modification'' of the radiation zone. Let us check the consistency of this assumption. 

The number of 3GeV-emitting electrons has been computed in \S\ref{S:tev} for the Vela pulsar: $N\sim 4.7\times 10^{29}$. The Goldreich-Julian number in the Vela is 
\be
N_{\rm GJ}\sim \frac{BR^2}{e}\sim 1.4\times 10^{31},
\ee
A factor $\sim 30$. The Vela efficiency is $\epsilon=1.3$\% -- a factor of $\sim 15$ below the mean weak-pulsar efficiency. Roughly a factor $\sim 10$ ``modification'' of the radiaton zone, as expected. This frivolous logic proves nothing, of course; a real numerical simulation of the pair production is highly desirable.

\section{Conclusion}

Tentatively, at the level of crude estimates, the recent discovery of 20 TeV radiation from the Vela pulsar confirms the Aristotelian Electrodynamics picture of the pulsar gamma-ray radiation: pulsars shine by annihilating colliding Poynting fluxes into curvature radiation near the light cylinder. In {\it weak} pulsars like the Geminga pulsar, this is all there is. In {\it strong} pulsars like the Vela pulsar, pair production modifies this processes, reducing the size and the strength of the radiation zone, but curvature radiation near the light cylinder is still happening and is responsible for the observe GeV photons.

In the Vela pulsar, according to the AE picture:
\begin{itemize}
\item The observed 3 GeV photons are the curvature radiation of electrons at about the light cylinder. 
\item Since the curvature radius is $R\sim 4\times 10^8$cm, the Lorentz factor is $\gamma \sim 5\times 10^7$, and the electron energy is $E\sim 30{\rm TeV}$. 
\item Given an acceptable target photon, the $30{\rm TeV}$ electron makes a $\sim 30{\rm TeV}$ photon by Compton scattering.
\end{itemize}


\begin{thebibliography}{99}

\bibitem{HESS} The H.E.S.S. Collaboration, Discovery of a Radiation Component from the Vela Pulsar Reaching 20 Teraelectronvolts, arXiv:2310.06181 (2023)

\bibitem{GJ} P. Goldreich, W.H. Julian, Pulsar Electrodynamics, Astrophysical Journal {\bf 157}, 869 (1969)

\bibitem{Fermi} The Fermi-LAT collaboration, The Second Fermi Large Area Telescope Catalog of Gamma-ray Pulsars, arXiv:1305.4385 (2013)

\bibitem{Mign} R. P. Mignani, R. Paladino, B. Rudak, A. Zajczyk, A. Corongiu, A. de Luca, W. Hummel, A. Possenti, U. Geppert, M. Burgay, G. Marconi, The first detection of a pulsar with the Atacama Large Millimetre Array, arXiv:1708.02828 (2017)

\bibitem{Ofen} D. D. Ofengeim, D. A. Zyuzin, Thermal Spectrum and Neutrino Cooling Rate of the Vela Pulsar, Particles, {\bf 1}, 194 (2018)

\bibitem{Fink} B. Finkbeiner, H. Herold, T. Ertl, H. Ruder, Effects of radiation damping on particle motion in pulsar vacuum fields, Astron. Astrophys. {\bf 225}, 479 (1989)

\bibitem{Gruz} A. Gruzinov, How Pulsars Shine: Poynting Flux Annihilation, arXiv:1402.1520 (2014)

\bibitem{Jaco} T. Jacobson, Structure of Aristotelian Electrodynamics, arXiv:1504.07311 (2015)

\bibitem{Gruzon} https://cosmo.nyu.edu/andrei/Papers/pulsars.html

\bibitem{Karg}  Kargaltsev, O. Y. ; Pavlov, G. G. ; Zavlin, V. E. ; Romani, R. W, Ultraviolet, X-Ray, and Optical Radiation from the Geminga Pulsar, arXiv:astro-ph/0502076 (2005)

\bibitem{Gruzs} A. Gruzinov, Aristotelian Electrodynamics solves the Pulsar: Lower Efficiency of Strong Pulsars, arXiv:1303.4094 (2013); Dissipative pulsar magnetospheres, arXiv:0804.4176 (2008)

\end{thebibliography}
\end{document}